\documentclass[conference]{IEEEtran}
\IEEEoverridecommandlockouts

\usepackage{tikz}
\usepackage{tikzscale}
\usepackage[mode=image]{standalone} 
\usepackage{pgfplots}
\usepgfplotslibrary{groupplots,dateplot} 
\usetikzlibrary{patterns,shapes.arrows}
\pgfplotsset{compat=newest}
\usepackage{xcolor,xspace}
\usepackage{physics}
\usepackage{amsfonts}
\usepackage{amsthm}
\usepackage[hidelinks]{hyperref}
\usepackage[capitalise]{cleveref}
\usepackage{flushend}

\newtheorem{innercustomgeneric}{\customgenericname}
\providecommand{\customgenericname}{}
\newcommand{\newcustomtheorem}[2]{%
  \newenvironment{#1}[1]
  {%
   \renewcommand\customgenericname{#2}%
   \renewcommand\theinnercustomgeneric{##1}%
   \innercustomgeneric
  }
  {\endinnercustomgeneric}
}

\newcommand{\ie}{i.\,e.\xspace}
\newcommand{\eg}{e.\,g.\xspace}

\newcommand{\mP}{\mathbb{P}}

\newcommand{\bC}{\boldsymbol{C}}

\newcommand{\rxx}{R_{xx}}
\newcommand{\ryy}{R_{yy}}
\newcommand{\rzz}{R_{zz}}
\newcommand{\rx}{R_{x}}
\newcommand{\ry}{R_{y}}
\newcommand{\rz}{R_{z}}

\newcommand{\CX}{\operatorname{CX}}
\newcommand{\ID}{\operatorname{I}}
\newcommand{\SX}{\sqrt{\operatorname{X}}}
\newcommand{\X}{\operatorname{X}}
\newcommand{\Phase}{P}

\newcustomtheorem{customprop}{Proposition}

\begin{document}
\title{Quantum Circuit Evolution on NISQ Devices}

\author{
Lukas~Franken,
Bogdan~Georgiev,
Sascha~Mücke, \\
Moritz~Wolter,
Raoul~Heese,
Christian~Bauckhage, 
and Nico~Piatkowski
\thanks{Lukas Franken is with the University of Edinburgh. Bogdan Georgiev is affiliated with Google DeepMind. Sascha Mücke is with the TU Dortmund University and Machine Learning Rhine-Ruhr (ML2R). 
Moritz Wolter and Christian Bauckhage are affiliated with the University of Bonn. Raoul Heese is member of the Fraunhofer ITWM, Kaiserslautern, Germany. Christian Bauckhage and Nico Piatkowski are members of the Fraunhofer IAIS, Sankt Augustin, Germany and Machine Learning Rhine-Ruhr (ML2R).  Moritz Wolter is a member of Fraunhofer SCAI. Lukas Franken and Bogdan Georgiev were affiliated with the Fraunhofer IAIS when working on this paper.
}}

\maketitle

\begin{abstract}
Variational quantum circuits build the foundation for various classes of quantum algorithms. 
In a nutshell, the weights of a parametrized quantum circuit are varied until the 
empirical sampling distribution of the circuit is sufficiently close to a desired outcome. 
Numerical first-order methods are applied frequently to fit the parameters of the circuit, 
but most of the time, the circuit itself, that is, the actual composition of gates, is fixed.
Methods for optimizing the circuit design jointly with the weights have been proposed, 
but empirical results are rather scarce. 
Here, we consider a simple evolutionary strategy that addresses the trade-off between finding appropriate circuit architectures and parameter tuning. We evaluate our method both via simulation and on actual quantum hardware. Our benchmark problems include the transverse field Ising Hamiltonian and the Sherrington-Kirkpatrick spin model. 
Despite the shortcomings of current noisy intermediate-scale quantum hardware, we find only a minor slowdown on actual quantum machines compared to simulations. Moreover, we investigate which mutation operations most significantly contribute to the optimization. The results provide intuition on how randomized search heuristics behave on actual quantum hardware and lay out a path for further refinement of evolutionary quantum gate circuits.
\end{abstract}

\begin{IEEEkeywords}
variational quantum circuits, structure learning, evolutionary computation
\end{IEEEkeywords}

\section{Introduction}
\label{sec:intro}
The current era of noisy-intermediate scale quantum computing (NISQ) \cite{Preskill2018}
allows us to get in touch with a technology that, one day, might outperform classical digital computers on useful tasks. Quantum algorithms exist whose theoretical runtime guarantees supersede those of their classical counterparts \cite{Nielsen2010}. 
However, the noise inherent to NISQ machines prevents the application of well-known quantum algorithms with proven speedups. Oppositely, variational quantum eigensolvers (VQE) \cite{McClean2016} are more robust and hence well suited to the available hardware. In VQE, one iteratively optimizes a set of parameters with respect to their performance on a given cost function. Applications include, among others, ground state approximation \cite{arute2020hartree,peruzzo2014variational}, simulation of imaginary-time evolution \cite{mcardle2019variational} and quantum machine learning \cite{farhi2018classification}. However, NISQ devices still suffer from limitations such as low circuit depth caused by large error probabilities and short decoherence times. Moreover, the recently exposed problem of barren plateaus \cite{McClean2018} causes gradients of cost functions to become exceedingly small as the number of system qubits is increased. 
In turn, this diminishes some of VQE's potential for problems of a practically relevant size \cite{kandala2017hardware}. To bypass such issues, we consider evolutionary strategies 
for learning the parameters of circuits, which removes the need for gradient computations and further allows us to estimate the circuit structure jointly with the parameters. 

\begin{figure}
\centering
\includestandalone{repcount}
\caption{Expected energy versus calls to a quantum computer for a gradient-descent method (GD) and our proposed evolutionary method (QNEAT).
The expected energy of an 8-qubit transverse field Ising Hamiltonian has to be minimized.
QNEAT performs a $(1+4)$ EA using the gates $\lbrace \rx, \ry, \rz, \rxx, \ryy, \rzz\rbrace$, starting with a circuit containing one random gate.
GD starts with a random Pauli 2-design ansatz with 40 parameters and computes derivatives via parameter shifts \cite{banchi2021measuring} used for parameter updates with fixed learning rate $\eta = 0.01$.
Five runs with random initial conditions yield mean and standard deviation.
The variance stems from the random initial gates and the particular mutations for QNEAT, and from the random ansatz and parameter initialization for GD.
The experiment was performed using a noise-free state vector simulation.}
\label{fig:repcount}
\end{figure}

\begin{figure*}
    \centering
    \includegraphics[width=.45\textwidth]{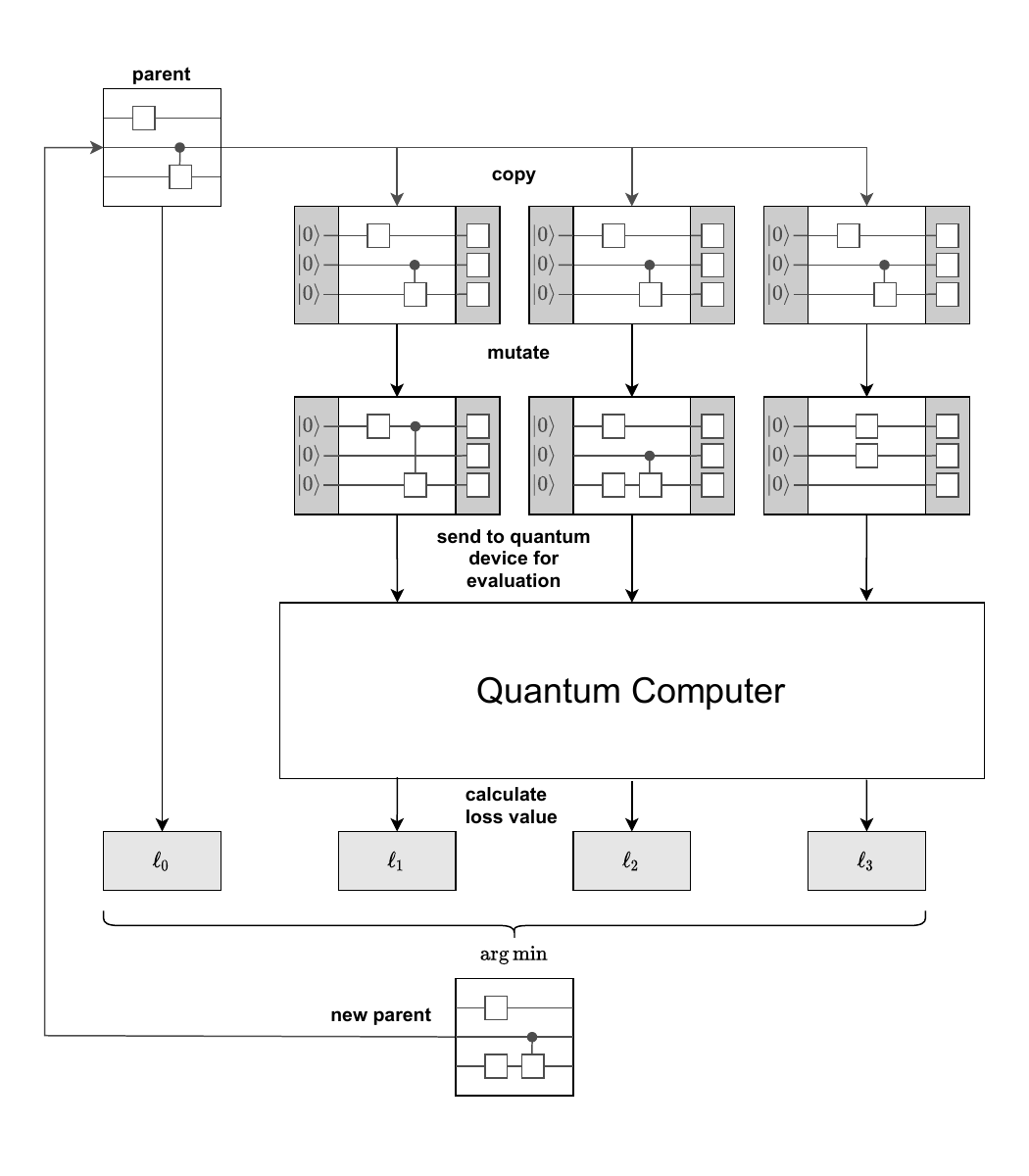}
    \includestandalone[scale=0.8]{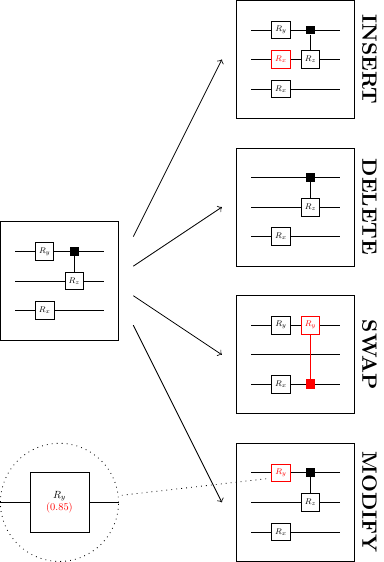}
    \caption{(Left) Outline of evolutionary circuit discovery:
    Starting with a random initial parent with a known loss value, we create copies and mutate them with our custom mutation strategy relying on operations shown on the right. The offspring population is then evaluated on the quantum machine.
    The circuit with the lowest loss value (the original parent included) becomes the parent circuit of the next generation.
    This procedure is repeated until an appropriate convergence criterion is satisfied. (Right) Mutation performs one of 4 actions with certain probabilities. We show these on the right.
    \textbf{INSERT:} Add random gate at a random position;
    \textbf{DELETE}: Delete gate at a random position;
    \textbf{SWAP}: Replace gate at a random position with a random new gate;
    \textbf{MODIFY}: Change parameter of a gate at a random position.
    }
    \label{fig:algorithm-outline}
\end{figure*}

\paragraph{Our contribution}
We provide a comprehensive empirical evaluation of evolutionary optimization of quantum gate circuits \cite{Nielsen2010} with respect to a goal function that corresponds to the expectation value of some target Hamiltonian. The mutations of our scheme feature insertion, deletion, swapping, and modification of circuit gates. The Hamiltonian of a quantum system is an operator (or matrix) corresponding to the total energy of that system. By testing the algorithm on Hamiltonians with varying difficulty, we show how the importance of these operations depends on the given problem. In case of a local Hamiltonian, the algorithm chooses to mostly forgo usage of the latter three operations and instead inserts gates in $\approx 7\%$ of all cases. Oppositely, for a Spin-Glass Hamiltonian, the success rate of the swapping operation is highest at about $11\%$, indicating that for more difficult problems gate insertion might be insufficient for successful optimization. These results refine upon the observations of \cite{tang2019qubit}. Also, we compare how our scheme's performance is reduced when run on actual quantum hardware. Consistent with expectation, optimization towards ground states prepared by non-local gates suffers non-trivial slowdown when run on IBMQ machines \emph{Manhattan}, \emph{Toronto} and \emph{Paris} \cite{ibmq2021}. Finally, we compare the speed of convergence to gradient descent methods and find the evolutionary scheme to drastically outperform the gradient method as sketched in \cref{fig:repcount}. However, we note that such performances are subject to a strong dependence on system size, and the current literature still lacks the respective theoretical treatment. We are nevertheless optimistic that evolutionary strategies retain comparable performance for larger quantum systems.

\paragraph{Related Work}
One of the first proposed variational quantum eigensolvers discussed trapped-ion computers for quantum chemistry \cite{Yung2014}. It represents the foundation of most existing techniques so far. 

Unfolding concurrently to variational approaches \cite{grimsley2019adaptive}, work at the intersection of evolutionary algorithms and quantum computing broadly falls into two categories: quantum-inspired evolutionary algorithms for classical computers and simulated quantum evolutionary algorithms.

Quantum-inspired evolutionary algorithms for classical computers simulate quantum bits, gates, superposition, and measurement to solve various problems within the usual standard framework \cite{zhang2011quantum}.
In general, this line of work intends to benefit from a richer quantum representation. Simulated quantum bits allow linear superposition of multiple states and are handled by synthetic quantum gates \cite{zhang2011quantum}. The concept is applicable to deep neural network architecture optimization, where it produces effective yet simple convolutional networks \cite{szwarcman2019Quantum}. However, computational costs of quantum simulations appear to be considerable. For example, the authors of \cite{szwarcman2019Quantum} report that 20 Nvidia K80 GPUs ($20 \cdot 24=480$ GB of GPU Ram) were required for two days.

Simulated quantum evolutionary algorithms seek to utilize evolutionary algorithms in a simulated quantum computation environment. Early work evolved a solution to Deutsch's problem \cite{spector1998genetic}. More recently, the Ising model of quantum computation was used to evolve multiple quantum gates in simulation \cite{krylov2019quantum}. Similarly \cite{mucke2019hardware} utilized
qubit encoding and the Ising model to create quantum-like behavior on FPGA-Hardware.

Quantum evolutionary computing has long been held back by limited availability and access of working quantum hardware~\cite{sofge2006toward,lahoz2016quantum}. Recently, however, the evolutionary approach has gained more traction by reducing the quantum computational overhead of exceedingly deep ansätze \cite{cerezo2020variational}. Particularly the Adapt-VQE algorithm \cite{grimsley2019adaptive} has shown promising results by alternating between optimizing ansatz and parameter configuration. In \cite{rattew2019domain} the authors investigate the potential of purely evolutionary-based algorithms to minimize ground Hamiltonian expectation value. Another issue potentially addressed by evolutionary algorithms are vanishing gradient phenomena \cite{anand2020natural}. However, other work expects the problem to persist in gradient-free optimization \cite{arrasmith2020effect}.
Finally, additional problems like abrupt training transitions can arise \cite{Campos/etal/2021a}.

\section{Notation and Background}
Let us summarize some necessary notation and background information used throughout this paper.

\subsection{Quantum Gate Circuits}
Quantum computation can be described by \emph{quantum circuits} operating on \emph{quantum states}.
In this context, an $n$-qubit quantum state $\ket{\psi}$ can be understood as a $2^n$-dimensional complex vector. States are always normalized such that $\braket{\psi}{\psi} = 1$, where $\braket{a}{b}$ denotes the ordinary inner product between states $\ket{a}$ and $\ket{b}$.
An $n$-qubit quantum circuit $\bC$ corresponds to a unitary operator (or $2^n \times 2^n$ matrix), which takes an input state $\ket{\psi_{in}}$---typically the all-$0$ state $\ket{0\cdots0} \equiv \ket{0}_1\otimes\ket{0}_2\otimes\dots\otimes\ket{0}_n$ with Kronecker product $\otimes$---and transforms it into an output state $\ket{\psi_{out}}=\bC\ket{\psi_{in}}$. Any unitary operator $U$ satisfies $U^\dagger U=U U^\dagger=\ID$ and its eigenvalues have modulus (absolute value) $1$. Here, $\ID$ denotes the identity and $U^\dagger$ represents the conjugate transpose of $U$.\par
To read out the computational results from $\ket{\psi_{out}}$, a \emph{measurement} $\ket{\psi_{out}} \mapsto \psi_{out}$ is performed that yields an $n$-bit vector $\psi_{out}\in\{0,1\}^n$. Such a measurement is inherently probabilistic. The squared inner product of a state $\ket{\psi_{out}}$ and a basis state represents the probability of measuring the corresponding bit string, \eg, the probability of measuring the bit string $10$ is $\vert\braket{10}{\psi_{out}}\vert^2$. After a measurement, the state $\ket{\psi_{out}}$ is destroyed and cannot be measured again. However, by performing multiple evaluations of identical circuits (each with a measurement), statistics like an expectation value can be obtained.
\par
The action of a quantum circuit can also be written as a product of unitary operators, \ie
\begin{equation}
	\bC = U_d U_{d-1} U_{d-2} \dots U_1,
\end{equation}
where $d$ is the \emph{depth} of the circuit.
Borrowing terminology from digital computing, the unitary operators $U_i$ are also called \emph{quantum gates}.
In theory and practice, they typically act on only one or two qubits at a time but can be composed via matrix multiplication and Kronecker products to form more complicated qubit transformations (since unitary operators are closed against matrix multiplication and Kronecker products).
It is important to emphasize that a quantum gate computer receives its circuit symbolically as a sequence of low dimensional unitaries---the implied $2^n \times 2^n$ matrix is never materialized.
\par
In the context of this work, we are particularly interested in the one-qubit unitaries
\begin{equation} \label{eqn:sigma}
\sigma^{(x)} \equiv \begin{bmatrix}
0 & 1\\
1 & 0
\end{bmatrix}
,\hspace{0.25cm}
\sigma^{(y)} \equiv \begin{bmatrix}
0 & -i\\
i & 0
\end{bmatrix}
,\hspace{0.25cm}
\sigma^{(z)} \equiv \begin{bmatrix}
1 & 0\\
0 & -1
\end{bmatrix}.
\end{equation}
Those matrices are also called \emph{Pauli matrices} and form the building blocks of the target Hamiltonians in the experimental section. A Hamiltonian $H$ can in this context be understood as a Hermitian operator (or $2^n \times 2^n$ matrix), \ie $H^\dagger=H$. As a consequence, all eigenvalues of $H$ are real. The eigenvalue spectrum describes the energy levels of a corresponding physical system. For a more detailed description of quantum gate circuits and related topics, we refer to \cite{Nielsen2010} and references therein. 

\begin{figure*}
\centering
\includegraphics[width=1.\textwidth]{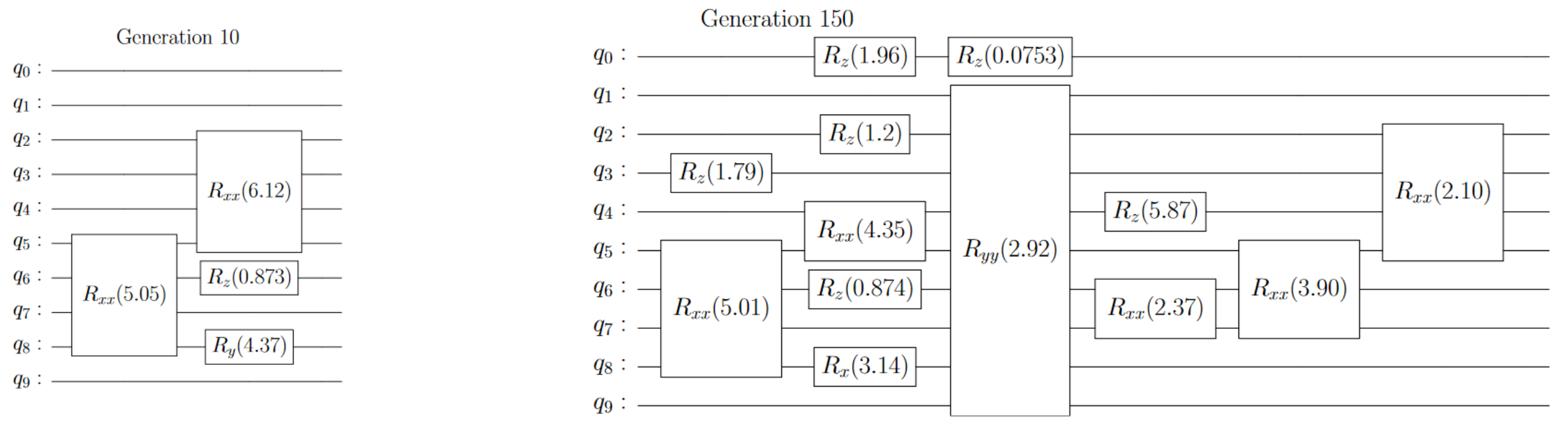}
\caption{Circuit diagram of the gradual evolution of the circuit architecture optimization the transverse field Ising Hamiltonian, (Left) after 10 generations and (Right) after 150 generations.
The gates represent rotations of one or two qubits as defined in \cref{eqn:R,eqn:RR}. For reasons of simplicity, only two digits of the rotation angles are shown. Note that gates spanning multiple qubits are only applied to the upper- and lowermost qubit. We remark that by generation 150 some previously present gates with redundant parameters are removed.}
\label{fig:arch}    
\end{figure*} 

\subsection{Evolutionary Circuit Learning}
In general, evolutionary algorithms (EA) \cite{fogel1994introduction, freitas2009review} iteratively work with a \emph{population} of candidate solutions 
and optimization is carried out over a number of \emph{generations}.
In each generation, $\mu$ candidates that constitute the \emph{parent} population produce an \emph{offspring} population of $\lambda$ candidates by means of crossover and mutation; these operations are specific to the problem domain at hand.
In our application, candidates are quantum circuits, and mutations cause small changes to the circuit, such as addition or removal of a gate or nudging of a gate's parameter.
The newly obtained offspring candidates are evaluated with respect to the loss function and sorted into the parent population, replacing parents with higher loss values.
This \emph{selection} step emulates natural selection in biology, where better-adapted individuals survive.
The parent population is maintained during selection. Consequently, the best individual that was ever observed since the first generation always survives.
This property is known as \emph{elitism} and ensures that the overall best loss value always monotonically decreases over time.
The $\mu$ best candidates proceed to form the parent population of the next generation.
If none of the offspring individuals yields an improvement, the original parent population carries over to the next generation unchanged.
This EA scheme is known in the literature as a $(\mu+\lambda)$ EA and is among the most representative and best-understood evolutionary strategies \cite{jansen2006analysis,droste2002analysis}.

The classical Neuro-Evolution of Augmenting Topologies (NEAT) algorithm \cite{Stanley2002} adapts evolutionary algorithms for learning the neural network structure jointly with weight optimization.
Evolution initially begins with a minimal structure to reduce the risk of evolving an overly complex solution.
Due to the competing conventions problem \cite{Stanley2002}, we choose to forego crossover. 
Our non-mating evolutionary algorithm evolves candidate solutions by a domain-specific mutation operator, which we describe in the upcoming section.

\section{Algorithm Outline}
\label{sec:outline-algorithm}
Motivated by the NEAT procedure, our method operates on quantum gate circuits instead of neural network structures and is therefore named \emph{QNEAT} algorithm. In particular, we define the required set of gates and the building blocks of the architecture. Possible solution circuits $\mathcal{U}$ should at least hypothetically be able to lay somewhere in the entire unitary group over the chosen number of qubits.
This is attainable by giving the algorithm access to a universal set of gates \cite{Nielsen2010}. Evolutionary optimization allows us to find a circuit $\bC \in \mathcal{U}$ that approximates the best possible circuit with respect to some cost function $f$.\par
The gates we consider are rotation gates
\begin{equation} \label{eqn:Uj}
    U_j \equiv U (g_j, \theta_j) \equiv \exp \left[ - i \, \frac{\theta_j}{2} \, g_j \right]
\end{equation}
defined by a unitary-parameter pair ($g_j$, $\theta_j$).
Here, the exponential function represents the matrix exponential.
The angles $\theta_j$ are from $[0,2\pi]$ and the unitaries $g_j$ come from the set $\{\sigma^{(x)},\sigma^{(y)},\sigma^{(z)},\sigma^{(x)}\otimes\sigma^{(x)},\sigma^{(y)}\otimes\sigma^{(y)},\sigma^{(z)}\otimes\sigma^{(z)}\}$, effectively generating the gate set $\mathcal{G} \equiv \{{\rx},{\ry},{\rz},{\rxx},{\ryy},{\rzz}\}$ consisting of one-qubit rotations
\begin{equation} \label{eqn:R}
	R_{n} \equiv R_{n}(\theta_j) \equiv U (\sigma^{(n)}, \theta_j)
\end{equation}
and two-qubit rotations
\begin{equation} \label{eqn:RR}
	R_{nn} \equiv R_{nn}(\theta_j) \equiv U (\sigma^{(n)} \otimes \sigma^{(n)}, \theta_j),
\end{equation}
where $n \in \{x,y,z\}$.\par
To verify that $\mathcal{G}$ is universal, we refer to a result from \cite{Barenco1995}, which proves that $\{{\ry},{\rz},\CX\}$ forms a universal gate set with the controlled NOT gate $\CX$ (which corresponds to a two-bit exclusive-or). One has $\CX = ({\ry}(-\pi/2)\otimes \ID) ({\rxx}(-\pi/2)) ({\rx}(\pi/2)\otimes{\rx}(-\pi/2)) ({\ry}(\pi/2)\otimes \ID) ( \Phase ( 7 \pi / 4 ) \otimes \Phase ( 7 \pi / 4 ) )$ with the global phase shift gate $\Phase( \delta ) \equiv \exp[ i \delta \ID ]$ and the identity gate $\ID = {\rz}(0)$, where ${\rx}(\delta) = {\ry}(\pi/2) {\rz}(\delta) {\ry}(-\pi/2)$ and $\delta \in [0,2 \pi]$. Since the global phase shift does not affect measurement outcomes, $\mathcal{G}$ is indeed universal.\par
The native gate set on IBMQ devices is in fact $\mathcal{G}' \equiv \{\ID,\rz,\X,\SX,\CX\}$, where $\X = {\rx}(\pi) \Phase(\pi/2)$ and $\SX = {\rx}(\pi/2) \Phase(-7\pi/4)$. In order to run a circuit consisting of gates from the set $\mathcal{G}$ on IBMQ hardware, the circuit is transformed into a representation of gates from $\mathcal{G}'$ in an automatic process called \emph{transpilation} \cite{ibmq2021transpiler}.\par
The cost value of a circuit is defined by the expectation with respect to some target Hamiltonian $H$ with $\ket{\psi_0}$ as initial state:
\begin{equation} \label{eq:target_function}
    f(\bC) \equiv \bra{\psi_0} \bC^\dag H \bC\ket{\psi_0}.
\end{equation}
Finding a circuit $\bC_{\mathrm{opt}}$ that minimizes $f$ is an optimization task over the search space $\mathcal{U}$.
This space $\mathcal{U}$ is defined through a mix of discrete and continuous values, namely unitaries $g_j$, qubit indices, and parameters $\theta_j$. 
While gradient methods are restricted to optimize the real-valued parameters $\theta_j$ on fixed circuit layouts, they cannot learn the overall circuit structure, since adding or removing gates are both non-differentiable operations.
Indeed, the gates from above are certainly differentiable around $\theta_j = 0$. Thus, insertion and deletion of gates can be simulated by gradient-based methods as well. However, this would require a very large initial circuit that basically contains all attainable circuits as a sub-structure---a method that has to be avoided due to the limited depth allowed for NISQ devices. 

However, evolutionary algorithms with elitist selection as described above can deal with non-differentiable and even non-continuous search spaces, because they only require some mutation operator $\mathfrak{m}$ that takes a circuit $\bC$ as input, applies random changes to it, and thus produces a slightly different circuit $\tilde{\bC}$.
More formally, $\bC$ is a random variable over $\mathcal{U}$.
If $\bC$ has support everywhere on $\mathcal{U}$, \ie $\mP(\bC=U)>0$ for all $U\in\mathcal{U}$, then the EA is guaranteed to converge to the global optimum \cite{rudolph1996convergence}.

Assuming $\mu=1$, we have a single parent circuit $U^{(t)}$ in generation $t$.
To find the parent of the following generation $t+1$, we sample $\lambda$ instances of $\mathfrak{m}(U^{(t)})$ and take their argmin with respect to $f$, including the parent $U^{(t)}$ itself:
\begin{align}
    U^{(t+1)} &= \underset{\tilde{U}\in\tilde{\mathcal{U}}}{\arg\min}  \sim f(\tilde{U}),\text{ where} \\
    \tilde{\mathcal{U}} &= \{\tilde{U}_i\sim\mathfrak{m}(U^{(t)}): 1\leq i\leq\lambda \} \cup\{U^{(t)}\}\nonumber
\end{align}

This process may be repeated until no more changes occur, \ie $f(U^{(t+\tau)})=f(U^{(t)})$ for a fixed threshold $\tau>0$, or some budget, such as a maximum number of computations on the quantum device, is depleted.

Note the distinct difference to the framework of regular VQE \cite{McClean2016} where a circuit is defined by a set of gates $U_i$, each of which carry a respective parameter $\theta_i$.
Then, a gradient method is used to optimize
\begin{equation}
  \theta_c = \underset{\vec{\theta}}{\arg\min} \: f(U(\vec{\theta})) \:,
\end{equation}
which is the setup where we encounter the caveats outlined in \cref{sec:intro}. 

The method described above is a combined evolutionary approach in the spirit of well-established network architecture search methods NEAT \cite{Stanley2002} thereby conducting a search over circuit architectures that intrinsically correspond to the problem. 
The macroscopic picture is shown in \cref{fig:algorithm-outline}. 
Each generation consists of two steps.
The specifics of the mutation strategy are explained in what follows.

\section{Experimental Evaluation}
We evaluate the performance of our QNEAT algorithm on standard VQE problems both in simulations and on actual quantum processors. 
For that purpose, we summarize the set of considered operations during mutation and outline our numerical approach.
We then describe physical systems with increasing difficulty, representing our benchmark problems on which we conduct the experimental evaluation in order to estimate the algorithm's general capabilities.\par

\subsection{Implementation and Method}
We use the publicly available quantum computing library qiskit \cite{aleksandrowicz2019qiskit} to define and modify quantum circuits in Python.
This library is used for interfacing with real quantum computers as well as simulation.

We use a custom EA implementation to perform an $(1+4)$-EA with a special multi-level mutation strategy.
We found the choice of $\lambda=4$ to be a suitable compromise between population diversity and speed.
The optimization run starts with a minimal random circuit, consisting of a single gate with a uniformly sampled parameter.
From this initial parent circuit, we make four copies and mutate them independently.
The resulting offspring circuits are sent to the IBMQ backend.
From the measurement results we derive a loss value, which we explain in detail in another paragraph further down this section.
The entire population is then sorted by loss value, and the circuit with lowest loss becomes the new parent for the next generation.
This process is repeated, with the parent's cost value monotonically decreasing, approaching the global optimum.

\paragraph{Mutation strategy}\label{par:mut}
The mutation strategy consists of a two-level random process.
Firstly, we choose an action from a list of options.
Secondly, we sample parameters for the chosen action and apply the action to the circuit at hand.
Possible actions (with their respective occurrence probabilities in parentheses) are:
\begin{itemize}
\item \textbf{INSERT} (50\%): Sample unitary $g$ and parameter $\theta$ uniformly and insert the corresponding gate at a random position.
\item \textbf{DELETE} (10\%): Delete gate at a random position from the circuit.
\item \textbf{SWAP} (10\%): Combination of \textsc{Delete} and \textsc{Insert} at the same randomly chosen position.
\item \textbf{MODIFY} (30\%): Modify parameter of randomly chosen gate according to $\theta\mapsto\theta+\epsilon$ with $\epsilon\sim\mathcal{N}(0,0.1)$.
\end{itemize}
The probabilities were found to perform best in a range of preliminary experiments.
See also \cref{fig:arch} for some visual examples.

With a probability of $0.1$, we repeat this entire mutation process after each action, leading to an expected number of $10/9=1.\overline{1}$ actions per mutation, the probability for 2 actions being about 9\%, for 3 actions about 0.9\% and for $k$ actions $0.1^{k-1}\cdot 0.9$ in general.
This scheme enables the mutation to perform arbitrarily large jumps in search space with positive probability, avoiding getting stuck in a local optimum indefinitely. 

\paragraph{Evaluation}\label{par:eval}
For the measurement process we append rotation layers to the circuit in accordance of the bases required by the Hamiltonian terms.
For the purpose of consistency, we leave most of the algorithm's hyperparameters, like population size and mutation action probabilities, unchanged for the majority of experiments.
The experiments are conducted for system sizes of 10 qubits.
In every generation, we collect thorough evolution data in order to extract information such loss value development over all generations, and types of mutations that lead to improvements.
This information is valuable for further analysis and improvement of the procedure.
\begin{figure}
\centering
\includestandalone[width=0.49\textwidth]{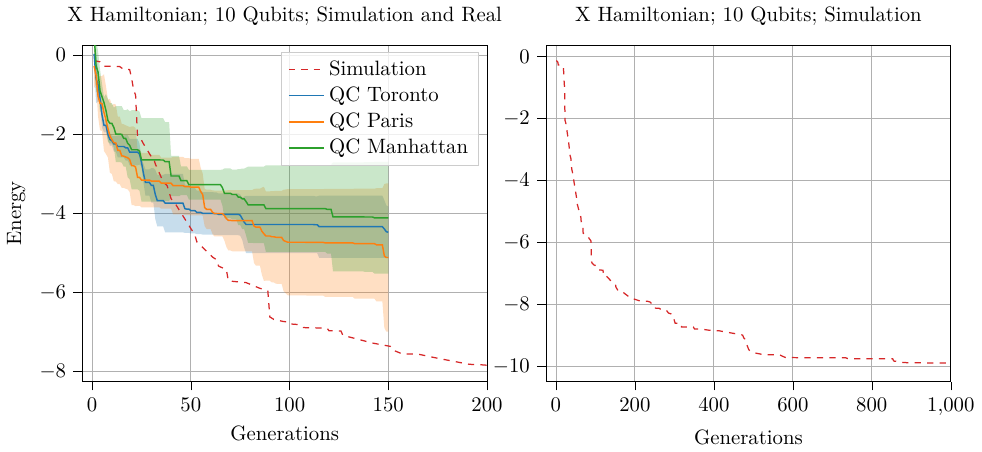}
\caption{Performance of the QNEAT algorithm on real quantum computers IBMQ Toronto, Paris, and Manhattan versus simulation for the maximally local problem on 10 qubits. All runs are shown as a mean of multiple runs with run length being limited by hardware availability. The exhibited performance is approximately similar between simulation and hardware, with a slight slowdown for the latter. Lower is better.}
\label{fig:X}    
\end{figure} 

\subsection{Considered Hamiltonians}
The difficulty to find the minimal eigenvalue to a Hamiltonian is intimately tied to the entanglement of its ground state.
Intuitively, the more entanglement required, the more intricate the optimization landscape is.
Therefore, we are first interested in our algorithms capability to perform in these non-convex circumstances.
Secondly, entangling qubits requires multi-qubit gates, which are much more error-prone on quantum hardware. 
This should be reflected in differences in performance between experiments on a simulator versus on an actual quantum machine.
To gradually raise the difficulty, we transition in three steps from a local problem to a spin-glass model only consisting of non-local terms.
This gives us a chance to test $a)$ our algorithm's capability to optimize increasingly difficult problems and $b)$ the effect of real-world machines on performance.
\paragraph{Local Hamiltonian}
As a sanity check we first consider a local problem
\begin{equation}
    \label{eq:local}
    H = \sum_{i=1}^n \sigma^{(x)}_i \: .
\end{equation}
Here $\sigma^{(x)}_i$ is the $x$-Pauli matrix acting on qubit $i$. 
Clearly, the ground state to this Hamiltonian is given by the $n$ qubit product state $\ket{\psi_0} = \ket{-}_1 \otimes \ket{-}_2 \otimes \dots \otimes \ket{-}_n$. 
For this relatively simple problem, we expect stable convergence in simulation and real hardware.
We start with local $\sigma^{(y)}$-eigenstate on all qubits.

\paragraph{Transverse Field Ising model (TFI)}
We next consider performance for the 1D spin-chain with correlation in the $z$-component and a transverse magnetic field with $x$-axis orientation \cite{pfeuty1970one}.
This scenario is captured by the Hamiltonian
\begin{equation}
    \label{eq:tfi}
    H = - J \sum_{\langle i, j \rangle}^n \sigma^{(z)}_i \sigma^{(z)}_j - h \sum_{i = 1}^n \sigma^{(x)}_i \: .
\end{equation}
We focus on the ordered phase and obtain (anti-) ferromagnetic behavior for ($J > 0$) $J < 0$. 
For our purposes, we chose $J, h = 1$, \ie opting for the anti-ferromagnetic behavior.
Note that for \cref{eq:tfi}, the entanglement is superimposed to the previous local problem, posing the question which part of the optimization the algorithm conducts first.
Our performance analysis also contains histograms of gates applied successfully during the optimization, giving us insights into the algorithm's choices at any stage.
The TFI model is known to exhibit local minima, causing purely gradient-based methods to fail \cite{wierichs2020avoiding}.
We expect our algorithm to be advantageous in this scenario.
For this problem we start with local $\sigma^{(y)}$-eigenstate on all qubits.

\paragraph{Sherrington-Kirkpatrick model (SK)}
The SK model simulates the behavior of a frustrated spin-glass \cite{sherrington1975solvable} and was previously used as a benchmark model in quantum computing experiments \cite{sung2020using}, \cite{arute2020hartree}. 
The model is given by the Hamiltonian
\begin{equation}
	\label{eq:sk}
    H = \sum_{i < j} J_{ij} \sigma^{(z)}_i \sigma^{(z)}_j,
\end{equation}
where $J_{ij}$ are randomly assigned couplings with $J_{ij} \in \{ -1, 1 \}$. 
For every run, we construct a randomly sampled instance of the SK model. 
However, the ground state energy of spin-glass model instances is subject to concentration hence properties such as optimization difficulty and ground-state entanglement can be expected to behave consistently \cite{ledoux2001concentration}. 
Note that such a Hamiltonian consists exclusively of correlation terms, which is why we expect this optimization problem to constitute a considerably harder problem than the previous experiments.
For this problem, we start with local $\sigma^{(x)}$-eigenstate on all qubits.

\subsection{Experiments}

\begin{figure}
\centering
\includestandalone[height=5.5cm]{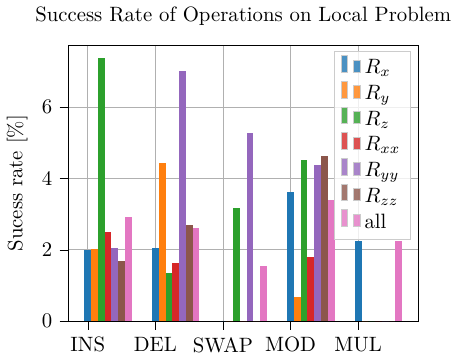}\vspace{.2cm}
\includestandalone[height=5.5cm]{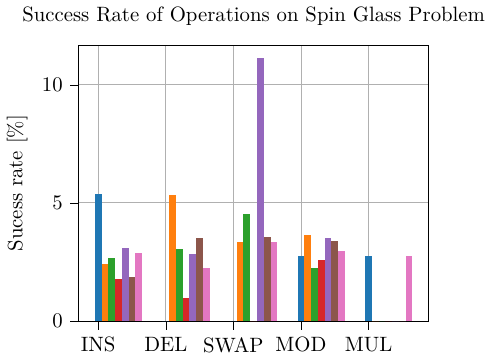}
\caption{Success rate of various mutation operations for the gates available to the circuit on (Top) the maximally local problem, \cref{eq:local}, and (Bottom) the SK Hamiltonian, \cref{eq:sk}. Each plot refers to a mean over all runs conducted with respect to that Hamiltonian (3 on each real quantum device). The operations on the $x$-axis are, from left to right, insert (INS), delete (DEL), swap and, modify (MOD). MUL indicates that a series of multiple operations was successful. We find all operations to provide a non-negligible number of useful additions to the optimization. Moreover, we observe distinct differences in the operations used between the Hamiltonians, where the trivial local problem exhibits a preference for gate insertion and the spin-glass problem for gate swapping.}
\label{fig:op_usage}    
\end{figure}

\begin{figure}
\centering
\includestandalone[width=0.49\textwidth]{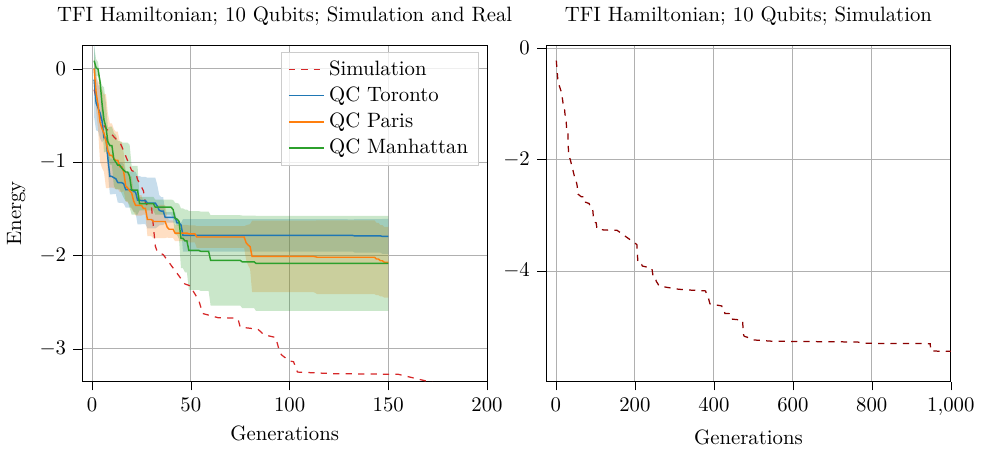}
\caption{Performance of the QNEAT algorithm on real quantum computers IBMQ Toronto, Paris and Manhattan versus simulation for the transverse field Ising model on 10 qubits. The initial phases of the algorithm show a larger similarity between simulation and real hardware compared to the optimization after generation 50. All runs are shown as a mean of multiple runs. For the optimization phase, exhibited performance was approximately similar. Lower is better.}
\label{fig:TFI}    
\end{figure}

\begin{figure}
\centering
\includestandalone[width=0.49\textwidth]{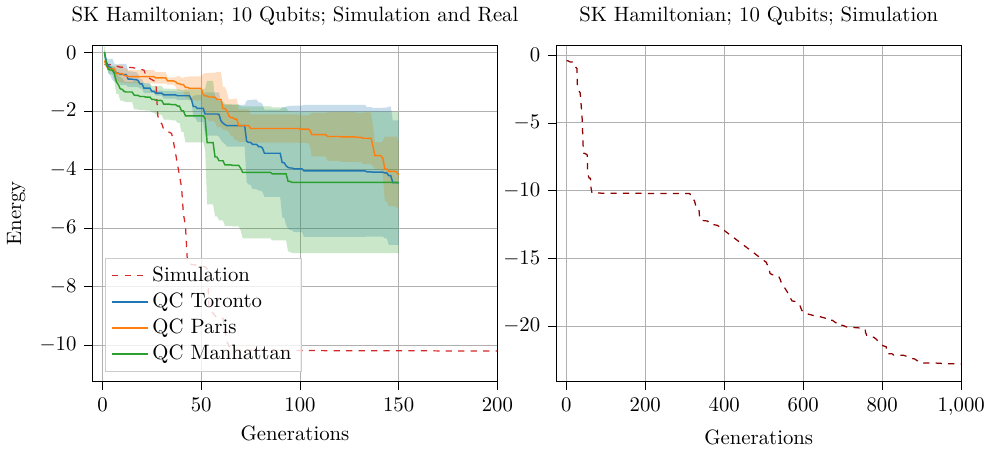}
\caption{Performance of the QNEAT algorithm on real quantum computers IBMQ Toronto, Paris and Manhattan versus simulation for the Sherrington-Kirkpatrick Hamiltonian on 10 qubits. All runs are shown as a mean of 5 runs where vertical lines indicate the ending of a run. Note that run length was limited by hardware availability. Simulation runs show reliable optimization with the runs on real hardware lagging behind by a substantial margin. In comparison to the TFI case, this slowdown is observable from the beginning to the run and not only after 50 generations. Lower is better.}
\label{fig:SK}    
\end{figure}

For our experiments, we used the IBMQ Manhattan, Toronto and Paris hardware backends alongside the qiskit state vector simulator.
All circuits had 10 qubits.
\cref{fig:TFI,fig:SK} show lowest energy per generation for each platform; the right-hand side plot shows only the simulation runs for more generations.
The energy is plotted as mean and variance of 5 evolution runs.
Additionally, to gain more insight into the algorithm's choices, we recorded which gates and evolutionary operations contributed positively to the optimization process, shown as a histogram in \cref{fig:op_usage}.
Overall, we find our algorithm to perform well on all posed problems dealing both with entanglement requirements and local minima. 
The histograms indicate intelligent circuit design both in gate and operation choice, mostly reflecting our intuitions about which rotations are useful for certain tasks.
We observe a slight reduction in performance on quantum hardware as soon as multi-qubit gates are required for optimization.

\paragraph{Results}

The results for the local problem are depicted in \cref{fig:X}.
In this experiment we optimize a trivial sum of local $\sigma_x$ Paulis over the full set of qubits. Clearly, the ground state then is given by the $n$ qubit product state $\ket{\psi_0} = \ket{-}_1 \otimes \ket{-}_2 \otimes \dots \otimes \ket{-}_n$ which can be prepared by local operations. 
Since the problem lacks a correlation requirement between the qubits we expect fast convergence for both the simulation and the real runs and observe so in \cref{fig:X} with all runs steadily converging to the global minimum.
The convergence is a bit steeper in the simulation, which is probably due to sampling noise when evaluating the Hamiltonian approximately from a fixed number of measurements, whereas the simulation has access to the true expectation value.
Since for this experiment we chose the $\sigma^{(y)}$ eigenstate on all qubits as the initial state, we expect rotations around the $z$-axis (in the plot denoted $\rz$) to be particularly useful.
This intuition is confirmed and we find that, by a large margin, this is indeed the rotation chosen most frequently according to \cref{fig:op_usage}.
Moreover, the figure shows that slight alterations to the chosen $z$-rotations or replacements in favor of $z$-rotations are preferred over gate deletions. 
Thereby, the algorithm admits to a gradient-like functionality once the appropriate set of gates is found.\par

We next considered performance for the transverse field Ising model, \cref{fig:TFI}.
The simulated runs show reliable convergence even for a low small number of offspring created per generation. 
Note that for TFI the optimization of the local problem is bounded by the number of qubits in the system hence where this threshold is surpassed, the algorithm optimizes also with respect to the entangling gates. 
The right-hand side clearly shows the desired behavior.
In the real hardware case, during the initial parts of the algorithm, we consistently observe optimization similar to the simulation yet notice a reduction in progress after $\sim$ 50 generations. 
Gate noise of deep candidate circuits might be a limiting factor here.\par

Finally, the results of optimization of the SK model are shown in \cref{fig:SK}.
In all cases, the early generations exhibit reliable optimization and show strong capability to navigate entangled Hilbert space.
However, the model's particular proneness to local minima is exhibited by longer optimization stretches without progress. 
These stretches are present particularly often in the simulation case, and we observe more stable optimization for the real hardware case.
Unfortunately, the required number of generations appears to be undercut quite substantially by the number of generations we were able to perform. 
However, \cref{fig:op_usage} (bottom) gives some insight into the algorithm's gate and operation preferences during SK optimization, exhibiting distinct differences to the previously discussed local problem. 
We find the algorithm to be more ``cautious'' in adding gates to the circuit and rather requires all gate choices to be finely tuned as can be seen from the relatively low frequency of gate insertions in comparison to gate swapping. 
For this problem we start with local $\sigma^{(x)}$-eigenstates in all qubits, making the $\sigma^{(y)}_i \sigma^{(y)}_j$ particularly useful.
Interestingly, such gates are predominantly added to the circuit via swapping and not via regular insertion. 
Whether this behavior is connected to the unusual Hilbert space traversal conducted by evolutionary algorithms in general, and the swapping operation in particular, is subject to further research.

\section{Conclusion}

We proposed an algorithm to alleviate issues common to variational quantum computing. During an iterative process, the method creates a set of mutants from a parent quantum circuit and subsequently determines the best performing candidates as parents for the next generation. We tested our algorithm on a set of Hamiltonians both as simulations and on actual quantum hardware.
On all posed problems, the algorithm reliably improves circuits according to the objective, with slight worsening for experiments on real-world quantum devices. Moreover, the algorithm admits to the following expected properties.\par

Non-locality impedes optimization, and as such, convergence on local problems proceeds faster. Between our experiments with exclusively local terms and the spin-glass model with only $zz$-correlation terms, this increase in hardness is reflected in our experimental results. 
Furthermore, the increased number of correlation terms in the Hamiltonian implies that more multi-qubit gates are required during the ground state preparation. Such gates are more affected by noise, which we observed in our experiments as a deceleration of runs on quantum hardware relative to simulation. Our statistics indicate that the circuit design is indeed intelligent, meaning not only are useful gates visibly preferred, but also are redundant gates removed over time. In comparison to gradient methods, this gradual circuit simplification is the most pronounced advantage over other VQE methods.\par

Comparisons to related work show that few approaches to gradient-free quantum optimization utilize the full spectrum of evolution strategies common in computer science literature. We show that all mutation operations (insertion, deletion, swapping, and modification) have approximately equivalent success rates, \ie rates at which the operations provide a useful alteration to the circuit. This emphasizes that evolutionary strategies applied to quantum optimization are most effective with the full spectrum of operations available to the algorithm.

\section*{Acknowledgement}
This work was supported by the Fraunhofer Cluster of Excellence Cognitive Internet Technologies (CCIT) and the Competence Center for Machine Learning Rhine-Ruhr (ML2R). ML2R is funded by the Federal Ministry of Education and Research of Germany (BMBF) (grants no. 01IS18038A and 01IS18038B).

\bibliographystyle{IEEEtran}

\end{document}